# Anomalous Raman Modes in Tellurides

F. J. Manjón,[o,*] S. Gallego-Parra,[o] P. Rodríguez-Hernández,[#] A. Muñoz,[#] C. Drasar,[$] V. Muñoz-Sanjosé,[&] and O. Oeckler[¥]

[o] Instituto de Diseño para la Fabricación y Producción Automatizada, MALTA Consolider Team, Universitat Politècnica de València, 46022 Valencia, Spain

[#] Departamento de Física, Instituto de Materiales y Nanotecnología, MALTA Consolider Team, Universidad de La Laguna, 38205 Tenerife, Spain

[$] Faculty of Chemical Technology, University of Pardubice, Pardubice 532 10, Czech Republic

[&] Departamento de Física Aplicada i Electromagnetismo, Universitat de València, 46100 Burjassot, Spain

[¥] Institut für Mineralogie, Kristallographie und Materialwissenschaft, Universitat Leipzig, Germany

AUTHOR INFORMATION

**Corresponding Author**

*fjmanjon@fis.upv.es

ABSTRACT

Two broad bands are usually found in the Raman spectrum of many Te-based chalcogenides, which include binary compounds, like ZnTe, CdTe, HgTe, GaTe, GeTe, SnTe, PbTe, $GeTe_2$, $As_2Te_3$, $Sb_2Te_3$, $Bi_2Te_3$, $NiTe_2$, $IrTe_2$, $TiTe_2$, as well as ternary compounds, like GaGeTe, $SnSb_2Te_4$, $SnBi_2Te_4$, and $GeSb_2Te_5$. Many different explanations have been proposed in the literature for the origin of these two anomalous broad bands in tellurides, usually located between 119 and 145 $cm^{-1}$. They have been attributed to the own sample, to oxidation, to the folding of Brillouin-edge modes onto the zone center, to the existence of a double resonance, like that of graphene, or to the formation of Te precipitates. In this paper, we provide arguments to demonstrate that such bands correspond to clusters or precipitates of trigonal Te in form of nanosize or microsize grains or layers that are segregated either inside or at the surface of the samples. Several mechanisms for Te segregation are discussed and sample heating caused by excessive laser power during Raman scattering measurements is emphasized. Finally, we show that anomalous Raman modes related to Se precipitates also occur in selenides, thus providing a general vision for a better characterization of selenides and tellurides by means of Raman scattering measurements and for a better understanding of chalcogenides in general.

**TOC GRAPHICS**

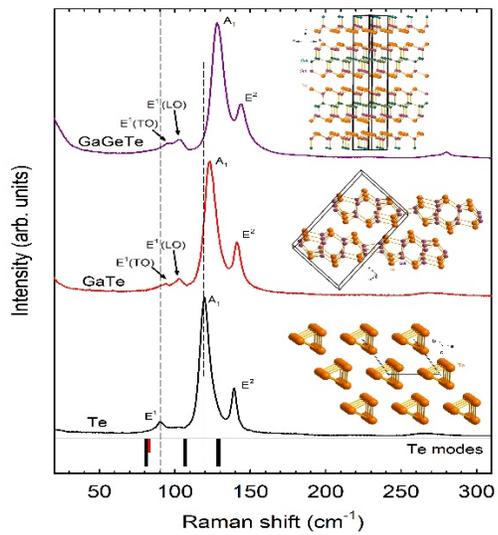

**KEYWORDS** Tellurides, Selenides, Raman scattering, laser heating, segregation

Since the boom of graphene, the study of 2D materials has increased exponentially and a strong interest has aroused in Te-based chalcogenides for photonic and optoelectronic applications. Consequently, a number of studies have been carried out on tellurides with different structure-types and compositions with special interest in van der Waals compounds. Among the common experimental techniques used to characterize bulk and 2D materials Raman scattering plays an important role since it can efficiently detect subtle structural changes due to atomic rearrangements in a non-destructive way that allows in situ characterization of materials and devices. Therefore, any consideration regarding the performance and common trends found in Raman spectra (RS) are of fundamental importance for a proper characterization of materials and devices, as was shown in the past, for instance, for ZnO.[1]

Among the vast literature concerning Raman scattering studies in Te-based chalcogenides one can find RS that are very similar to many tellurides despite their different compositions and even crystalline structures. These common Raman features found in different tellurides are hereafter named anomalous Raman modes (ARMs) and give rise to anomalous Raman spectra (ARS). In order to shed light on the origin of the ARMs in tellurides we have performed a joint experimental and theoretical study. For this purpose, Raman-active mode frequencies obtained from Raman scattering measurements performed in several bulk tellurides have been compared to ab initio theoretical simulations (see experimental and theoretical details in the Supplementary Information (SI)). Our study concludes that the ARS common to many Te-based chalcogenides come from Te clusters or precipitates that eventually could dominate the Raman spectrum, especially in nanometric 2D tellurides. Moreover, our study also provides proofs that ARMs coming from Se precipitates are observed in selenides.

In order to approach the problem, we show in **Figure 1** the RS of monoclinic GaTe and rhombohedral GaGeTe. **Figures 1a and 1e** show good agreement with those early reported in the literature[2-6] and with our theoretically calculated wavenumbers for the first-order Raman-active modes in both compounds (see tick marks in **Figure 1**). However, another kind of RS can be found in these samples (**Figures 1c and 1g**) or even a mixture of them (**Figures 1b and 1f**) in different zones of the same samples. As it can be observed, the RS of these zones of the samples look rather similar despite the different composition and crystalline structure of both compounds. The Raman bands shown in these RS are considered to be the ARMs in tellurides, so they are considered to be ARS. The ARS in the two compounds show two intense and broad bands (one in the 122-128 $cm^{-1}$ range and the other in the 141-144 $cm^{-1}$ range). In fact, ARS measured on several regions of both samples evidence that these bands do not always have the same central wavenumbers. The first band is usually located between 119 and 130 $cm^{-1}$ and the second one is usually found between 139 and 145 $cm^{-1}$ (see **Figures 1b, 1c, 1f and 1g**). In any case, the first band is always more intense and broader than the second one. In particular, the linewidth of those bands, defined by the full width at half maximum (FWHM), is found to be around 9.5 and 8 $cm^{-1}$, respectively, in **Figures 1c and 1g**. A more detailed view of the ARS shows several weaker bands near 100 $cm^{-1}$ and between 260 and 300 $cm^{-1}$ that can be also considered as ARMs.

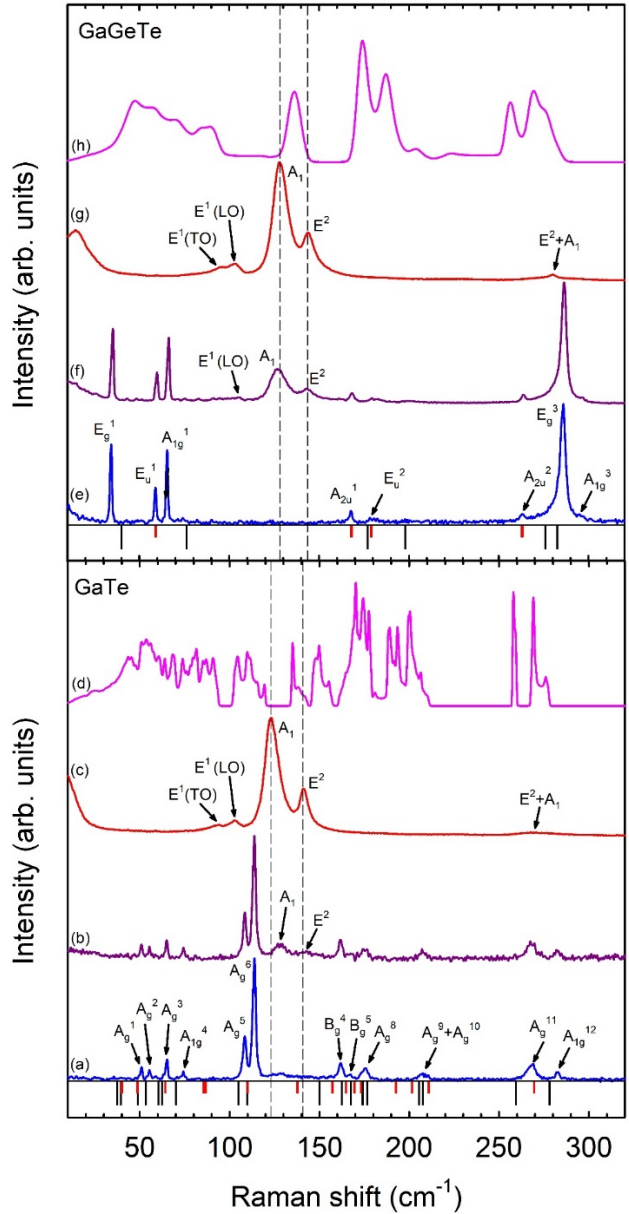

**Figure 1.** Normalized and unpolarized Raman spectra of monoclinic GaTe and rhombohedral GaGeTe in different zones of the samples. (a) Normal Raman spectrum of GaTe. (b) Raman spectrum of GaTe with some anomalous bands. (c) Raman spectrum of GaTe showing only anomalous bands. (d) One-phonon density of states of GaTe. (e) Normal Raman spectrum of GaGeTe. (d) Raman spectrum of GaGeTe with some anomalous bands. (f) Raman spectrum of GaGeTe showing only anomalous bands. (g) One-phonon density of states of GaGeTe. (h) Optical microscope image of the region of GaGeTe with good Raman signal. Bottom black (red) tick marks show the calculated Raman-active (IR-active) TO modes of GaTe and GaGeTe. Dashed lines show the positions of the anomalous Raman bands as well as the main peak of GaTe and GaGeTe. Spectra have been normalized and vertically shifted for the sake of comparison and clarity.

The most striking point is that most of the bands observed in the ARS of GaTe and GaGeTe, especially the two strongest ones, have been observed in a number of 2D and bulk Te-based chalcogenides with different laser wavelenghts (typically with green and red lasers), irrespective of their different compositions and crystalline structures. They have been observed in ZnTe,[7,8] CdTe,[9-14] GaTe,[15-22] $As_2Te_3$,[23] $Sb_2Te_3$,[24,25] $Bi_2Te_3$,[26,27] GeTe,[28,29] SnTe and PbTe,[30,31] $GeTe_2$,[32-35] $TiTe_2$,[36-39] GaGeTe,[6] $SnSb_2Te_4$,[40] and crystalline $Ge_2Sb_2Te_5$,[41] to name a few. A search among different tellurides shows that in most cases the first and more intense ARM is observed between 119 and 130 cm$^{-1}$, while the second ARM is usually observed between 139 and 145 cm$^{-1}$. Several questions arise: Which is the origin of the ARS in tellurides? Can they be attributed to the same origin in all tellurides? How can they be formed? Why are ARMs so prominent in many RS so as to hidden the normal Raman modes of the corresponding compounds?

To give answer to these questions we have gone through the literature and found that there is an ongoing controversy. Regarding the nature of the ARMs, many authors generally attribute them to the own sample or simply to oxidation. In the first explanation, they have been attributed either to Raman-active modes of the sample or to IR-active modes observed in Raman scattering measurements due to the breakdown of Raman selection rules. However, this explanation cannot give account for these modes neither in GaTe nor in GaGeTe (see theoretically predicted Raman- and IR-active modes in both compounds as tick marks in **Fig. 1**). In monoclinic GaTe, IR-active modes are not observed in the RS and the calculated IR-active modes have wavenumbers that do not match with those of the two ARMs. Moreover, theoretically calculated Raman-active modes for the other known polymorph of GaTe, hexagonal GaTe, either in monolayer or bulk form[42,1] do not match with the ARMs. On the contrary, many IR-active modes are observed in the RS of GaGeTe[43] and match with the theoretically calculated values. Therefore, we can conclude that this explanation for the origin of the ARMs is not consistent; similar Raman- or IR-active modes cannot be observed in all tellurides with so different compositions and crystalline structures. Consequently, a different origin must be invoked for the ARMs in all tellurides.

Surface oxidation has also been proposed in many papers to explain the origin of the ARMs in tellurides. In some works, ARMs have been specifically attributed to the formation of $TeO_2$ layers. However, the three known polymorphs of $TeO_2$ at ambient conditions show narrow and intense Raman bands in a wide wavenumber region, with several strong peaks below 250 cm$^{-1}$, near 400 cm$^{-1}$ and above 600 cm$^{-1}$.[44-46] Consequently, the Raman modes of the $TeO_2$ polymorphs are not consistent with the observed and reported ARMs in tellurides,[30,31] as already noted in works that confirmed the presence of $TeO_2$ surface layers by X-ray Photoelectron Spectroscopy.[30] It must be noted that the 62 cm$^{-1}$ mode characteristic of paratellurite, the most stable phase of $TeO_2$, was observed in the RS of supposedly amorphous Te (latter attributed to $TeO_2$) obtained from melting due to the use of a relatively high laser power (125 mW) during Raman scattering measurements of pure trigonal Te.[44,47] However, such a Raman mode has not been observed on a regular basis neither in any of the ARS of tellurides already commented nor in our RS of GaTe and GaGeTe. Therefore, the formation of $TeO_2$ layers in tellurides cannot be the origin of the ARMs in tellurides.

More recently, molecular oxygen adsorbed in the sample surface or in the first atomic layers due to sample oxidation have also been proposed as the origin of the ARMs in GaTe films.[18] In particular, vibrational modes of GaTe-$O_2$ have been suggested as the cause for the two broad Raman bands. It is unlikely that the same oxygen molecules show the same ARS in all compounds with different compositions and, more importantly, crystalline structures. We must note that among tellurides there are many layered van der Waals-type compounds with Te-terminating layers but also non-layered compounds. In this context, one can find layered compounds showing flat layers (GeTe, $Sb_2Te_3$, GaGeTe, $SnSb_2Te_4$) irregular layers (GaTe) and zigzag layers ($\alpha$-$As_2Te_3$), as well as non-layered compounds with zincblende-like structure (ZnTe, CdTe, HgTe) and with rocksalt-type structure (SnTe, PbTe). Thus, it is unlikely that modes of GaTe-$O_2$, with $O_2$ molecules between the layers or at the surface, can equally give account for the ARMs in all layered and non-layered tellurides.

In some recent papers, the ARMs in tellurides have been attributed to other causes. In a study of GaTe films,[15] the two main ARMs have been attributed to second-order Raman scattering due to the existence of a double resonance in GaTe, like that of $MoTe_2$ and graphene.[48,49] The large linewidth and small polarization dependence of the two main ARMs in tellurides was claimed to give support to the hypothesis of the double resonance in GaTe.[15] In fact, **Figure S1** in the SI shows polarized and unpolarized Raman scattering measurements in GaGeTe that evidence that both ARMs show similar dependence on polarization, as already reported in the literature. However, it is unlikely that the same double resonance mechanism (likely valid for $MoTe_2$ but which do not show the two main ARMs of tellurides in **Ref. 48**) is also valid for ZnTe, CdTe, HgTe, GeTe, SnTe, PbTe, GaTe, $GeTe_2$, GaGeTe, and so on, whose crystalline structures and electronic band structures are completely different among them. We want to stress that while this hypothesis could give account for some second-order Raman modes in tellurides, like $MoTe_2$ with a bandgap around 1.1 eV,[48] it is rather unlikely that it can show similar resonances for GaTe, a semiconductor with a bandgap around 1.65 eV,[18] and GaGeTe, a semimetal with very small direct and indirect bandgaps.[50] Moreover, it is very unlikely that the resonances could occur by excitation with different laser wavelenghts in so different tellurides as those already commented. The theoretical electronic band structure of monoclinic GaTe and rhombohedral GaGeTe can be viewed in the Materials Project Database,[51,52] and their completely different nature can be checked, what makes improbable the observation of double resonance in both compounds with similar features as those shown in **Figure 1**. Therefore, the double resonance mechanism cannot be the explanation of the ARMs in so many tellurides and another explanation is needed.

A Raman-active mode originated from a Brillouin-edge mode of the M-point folded into the Brillouin-center ($\Gamma$-point) has also been recently proposed as the origin of the ARMs in TiTe$_2$.[38] Again, it is very unlikely that this explanation could be also valid for so different compounds, with completely different compositions and crystalline structures, thus giving completely different vibrational branches along the whole Brillouin zone. **Figure S2** in the SI shows the phonon dispersion curves of GaTe and GaGeTe. It can be observed that they are completely different and cannot yield similar ARS even by folding different Brillouin-edge points into the $\Gamma$-point. Consequently, another explanation for the ARMs in tellurides is required.

Defects or disorder have also been claimed as the origin of the ARMs in GaTe films.[16,17] It is well known that defects or disorder result in RS showing local vibrational modes in addition to those of the corresponding material or being dominated either by the one-phonon density of states or by the observation of silent modes.[1] Regarding disorder, it can be seen that the one-phonon density of states of monoclinic GaTe and rhombohedral GaGeTe (**Figures 1d and 1h**) do not agree with the ARMs observed in any of both compounds. In addition, there are no silent modes in GaTe and GaGeTe. On the other hand, regarding defects or impurities, they can give rise to local vibrational modes, which usually are relatively narrow bands, unlike the broad bands shown in the ARS. Consequently, the disorder or defect origin of the ARMs can be discarded. Note that this hypothesis could be a possible explanation for the ARMs in some compounds, but again it does not provide an explanation for the origin of the ARMs in other compounds since completely different local vibrational modes, silent modes and one-phonon density of states will be obtained for different tellurides with different composition and crystalline structure.

Finally, there are some works in which a much simpler explanation is given for the origin of the ARMs in tellurides. They have been tentatively attributed to the presence of defects in the form of Te precipitates; i.e. grains or layers of pure trigonal Te segregated from the original sample.[7-9,11,14,23,30,40] In order to substantiate this hypothesis, we have compared the ARS in **Figure 1** with the unpolarized RS (**Figure 2**) and polarized RS of crystalline trigonal Te for different polarizations (see **Figure S3** in the SI).

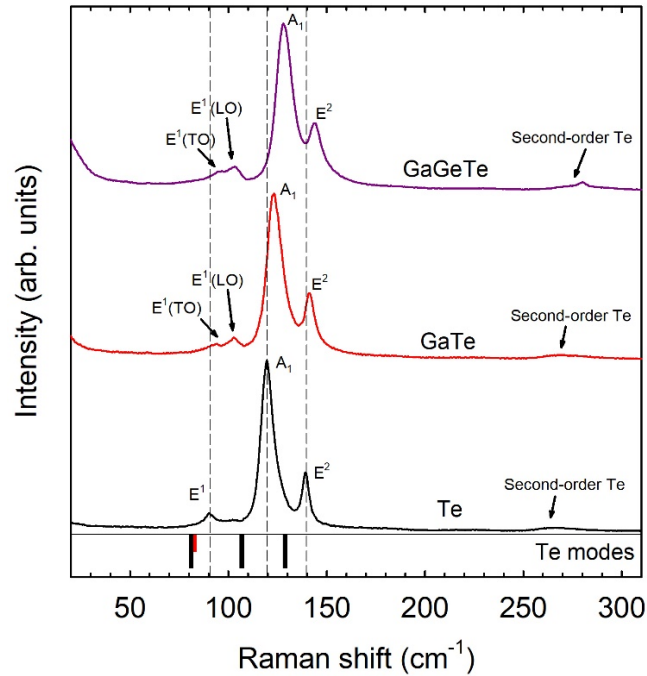

**Figure 2.** Comparison of the unpolarized Raman spectra of Te and the anomalous Raman spectra in GaTe and GaGeTe. Bottom black tick marks show the calculated Raman-active TO modes of trigonal Te ($E^1$, $A_1$, $E^2$). Bottom red tick marks show the calculated IR-active $A_2$(TO) mode of trigonal Te. Note that $E^1$ and $E^2$ modes of Te are also IR-active so $A_2$, $E^1$ and $E^2$ modes can also show LO features. Spectra have been normalized and vertically shifted for the sake of comparison and clarity.

As regards our unpolarized and polarized RS of crystalline Te, they are similar to those reported in the literature[9,47,53-55] and show three main bands around 91, 120 and 140 cm$^{-1}$ corresponding to the three first-order Raman modes $E^1$(TO), $A_1$ and $E^2$(TO) modes of trigonal Te. We have noted the two E modes of trigonal Te with superindexes 1 and 2 to distinguish for the low and high-wavenumber modes, respectively. Additionally, several weak bands attributed to second-order modes of trigonal Te have been observed in good agreement with the literature.[47,56-58] The FWHM of the three main bands of trigonal Te are: 3.2(4), 4.8(2) and 2.8(3) cm$^{-1}$ for the $E^1$, $A_1$ and $E^2$ modes, respectively; i.e. the $A_1$ mode shows larger linewidth than the two E modes. It can be also observed that the RS of trigonal Te are rather sensitive to polarization when the laser polarization (E) is either parallel or perpendicular to the *c* axis; however, the Raman modes of Te show much smaller sensibility to polarization of collected scattered light when laser polarization is at 45º with respect to the *c* axis.

It can be observed the strong similarity of our RS for pure trigonal Te with the ARS in GaTe and GaGeTe (**Figure 2**). All RS show two intense bands close to 120 and 140 cm$^{-1}$, weaker bands near 100 cm$^{-1}$ and much weaker bands near 260 cm$^{-1}$. Therefore, all the ARMs in these tellurides can be attributed to the first-order modes ($A_1$, $E^1$(TO), $E^1$(LO), $E^2$(TO)) and second-order modes of trigonal Te. A close comparison between them yields that most ARMs modes in tellurides are shifted to larger wavenumbers and are broader than those in pure trigonal Te. Additionally, ARS in GaTe and GaGeTe exhibit one extra peak near 100 cm$^{-1}$ and no sensibility to polarization (**Figure S1**). In summary, ARMs in tellurides show strong similarities with those in pure trigonal Te that suggest that they could come from Te precipitates, but they also show some differences that must be explained to further give support to this hypothesis.

In order to explain the blueshift of the wavenumbers of Te modes in the ARS of tellurides, we have commented that the two main ARMs in Te-based chalcogenides occur at different wavenumbers around 119-130 cm$^{-1}$ and 139-145 cm$^{-1}$. It has been shown that in general Raman modes of trigonal Te shift to higher frequencies in 2D Te as the number of tellurene layers decrease.[59-62] In fact, values of $A_1$ and $E^2$ modes as high as 136 and 149 cm$^{-1}$, respectively, have been measured in Te monolayers.[60] Only in one work, Raman peaks were found to shift first to larger wavenumbers and then to smaller wavenumbers in 2D Te as the number of layers decrease.[63] We think that this strange Raman shift behavior is likely due the compressive stress on Te layers below a certain thickness. Therefore, we consider that the reason for the larger wavelenghts observed in the ARMs of tellurides with respect to bulk Te can be ascribed to the small layer thickness or grain size of Te precipitates (of the order of nm to μm) in form of polycrystalline Te, as already observed in some earlier works of CdTe.[30] An estimation of the layer thickness or grain size of Te precipitates can be done on the basis of recent works.[59-62] According to these works, flakes of pure Te between 10 and 30 nm showed the two bands around 124-125 and 143 cm$^{-1}$, respectively, while flakes with thickness below 1 nm show bands at wavelenghts above 129 and 147 cm$^{-1}$, respectively. Therefore, it can be concluded that ARMs in tellurides with similar or larger wavenumbers than those here summarized can be considered to correspond to pure Te grains or layers with nanometric size.

It is interesting to note that while the hardening of the E-type modes are expected in 2D materials in comparison to bulk materials, the large hardening of the A-type mode of Te in 2D tellurium (contrary to that observed in other 2D materials) is still not fully understood in the context of standard covalent and van der Waals forces of layered materials.[60]

The larger linewidth of the ARMs in tellurides than in pure trigonal Te can also likely be ascribed to the nanometric nature of the Te precipitates in tellurides. Note that Te nanoprecipitates (grains of different size or layers with different thickness) will give rise to different wavenumbers for each Raman-active mode and to broader linewidths due to the relaxation of Raman selection rules. Consequently, RS of nanoprecipitates will result in the sum (convolution) of the Raman-active modes of Te precipitates with different grain sizes or layer thicknesses thus resulting in much broader linewidths than in bulk trigonal Te. It must be also mentioned that shift of the wavenumbers and broadening of ARMs in tellurides could be also partially attributed to strain in the segregated Te at the surface, as already suggested to occur in CdTe,[9] but we think that this shift will be a minor component in comparison to the shift caused by the nanometric nature of the precipitates.

Further support for the assignment of the ARMs in tellurides to Te precipitates is that both $E^1$(TO) and $E^1$(LO) modes of Te are observed in the RS of both GaGeTe and GaTe. The last one has a wavenumber around 104 cm$^{-1}$ in good agreement with previous works of trigonal Te.[9,54,55] The observation of the $E^1$(LO) mode of Te has been observed in 2D Te layers[60] and is likely due to the partial breakdown of the Raman selection rules in nanocrystalline-size Te precipitates. The lack of long range order in nanocrystalline-size grains allows the observation of the IR-active modes in the RS.[64] In fact, the contribution of IR-active $A_2$ modes of trigonal Te to the broad band near 100 cm$^{-1}$ in the ARS of GaTe and GaGeTe cannot be discarded.[54,55] The observation of the LO modes not only evidences the presence of the $E^1$(LO) mode but it also can contribute to the broadening of the $E^2$ mode since both $E^2$(TO) and $E^2$(LO) modes in trigonal Te show very similar wavelenghts, unlike for the $A_2$ and $E^1$ modes.[47,54,55] On top of that, the observation of the second-order Raman modes of Te above 260 cm$^{-1}$, as have been observed in the ARS of GaTe and GaGeTe, gives a fundamental support to their assignment to Te precipitates. These modes have also been reported in some tellurides.[18,22,34] In fact, a shift to larger wavelenghts is observed for the second-order modes above 260 cm$^{-1}$ in the ARS of tellurides compared to pure trigonal Te (**Figure 2**).

Finally, the small polarization sensibility observed in the ARMs in Te-based chalcogenides can also be ascribed to the formation of polycrystalline Te precipitates of very small and different grain sizes. All those small grains are randomly oriented with respect to the polarized laser light so they give rise to completely depolarized RS. Noteworthy, a completely depolarized RS was assumed to be related to initially assumed amorphous Te obtained by laser melting and recrystallization of pure Te,[47] that was latter attributed to paratellurite despite only one peak of paratellurite at 62 cm$^{-1}$ was clearly found[44] in comparison with what was reported in the literature.[44-46] We think that laser melting of pure trigonal Te reported in **Ref. 47** resulted in Raman modes of 62, 120 and 146 cm$^{-1}$ related to the formation of amorphous Te (Raman modes of 120 and 146 cm$^{-1}$) as well as of paratellurite (Raman mode of 62 cm$^{-1}$) due to strong surface oxidation favored by the high temperature reached upon excitation with 125 mW of laser power.

Several additional considerations support the assignment of the ARMs of tellurides to mainly nanocrystalline-size Te clusters or precipitates: i) The two strongest modes of Te ($A_1$ and $E^2$) always appear as a pair in most ARS in tellurides. ii) Always the intensity ratio of both ARMs is similar to that found in trigonal Te. iii) The FWHM of the first ARM is always larger than that of the second one as in trigonal Te and the FWHM of the two ARMs in Te precipitates is always larger than in pure Te as expected for layers of nanometric size. iv) The two strongest ARMs measured in some tellurides, like ZnTe,[7] CdTe,[7] α-$As_2Te_3$,[23] and $SnSb_2Te_4$,[40] show negative pressure coefficients as those reported for bulk trigonal Te.[65-67] Moreover, the first anomalous Raman mode shows a larger negative pressure coefficient than the second one, as in trigonal

Te.[7,23,40] The negative pressure coefficients shown by the ARMs are clear fingerprints of the Te nature of the ARMs in tellurides. Therefore, all the above mentioned features clearly indicate that the ARMs in tellurides are related to the formation of nanocrystalline trigonal Te either in form of small grains or layers. Since Te is common to all tellurides, Te segregation is a very reasonable hypothesis to explain the ARMs in so different tellurides as those here commented. These segregates could be either in the interior of the material or at the surface of the samples.

Two questions still to be answered are why ARMs from trigonal Te clusters at the sample surface are so strong and how can they be observed even when $TeO_2$ films over Te layers have been found in several compounds. To answer these questions we have to consider at least four factors: i) $TeO_2$ polymorphs are insulating phases with large bandgaps (well above 2.5 eV),[68] so excitation of inner layers of Te below the $TeO_2$ layers is always possible because of the large penetration length of visible light in $TeO_2$. ii) On the contrary, Te is a semimetal with a bulk bandgap around 0.33 eV and nanolayers have a much larger bandgap that can range from 0.65 to 1.17 eV.[69,70] With such small bandgaps for Te layers, excitation with visible light leads to a very small penetration depth as small as 500 Å (100 atomic layers).[71] Consequently, it is difficult to perform RS measurements of samples covered by a relatively thick Te layer, especially if the Raman signal of the sample is much smaller than that of trigonal Te. Noteworthy, the formation of Te layers on top or inside GaTe layers could explain not only the ARMs in the RS but also the apparent decrease of the bandgap of GaTe (from 1.65 eV to 0.77 eV) when exposed to air.[18] iii) It must be stressed that there is a strong resonance of Te Raman modes when excited with red and green laser lines that are usually employed for Raman scattering measurements in most laboratories.[54] Therefore, Te nanolayers have a fairly large Raman scattering cross section when excited with visible light that can avoid the observation of the Raman signal of samples below them. iv) Much larger Raman scattering cross section was measured for the supposed amorphous Te than for bulk crystalline Te.[47] Therefore, RS of Te thin films at the surface of samples show a strong signal that can even obscure the intrinsic Raman modes of the compound, as shown in the ARS of GaTe and GaGeTe in **Figure 1**. In summary, all these factors can explain why ARMs are observed in many tellurides instead of the expected Raman modes of the corresponding tellurides or in combination with them. A notable example are the recently RS measured in $TiTe_2$[36-39] that are rather different to those previously reported[72] and surprisingly similar to those of trigonal Te, including the negative pressure coefficient of the two strongest Raman modes of Te.[38,39]

At this point, we must stress that our explanation for the origin of the ARMs in Te-based chalcogenides is also valid for Se-based chalcogenides since ARMs due to Se segregation have also been observed in ZnSe,[7] CdSe,[14] $HgGa_2Se_4$,[73] $TiSe_2$,[74] $TaSe_2$[75] and $In_2Se_3$.[76] The ARMs in several of these selenides have been observed near 150 cm$^{-1}$ (corresponding to the $E^1$ mode of trigonal Se) and/or near 235 cm$^{-1}$ ($A_1$ and $E^2$ modes of trigonal Se are sometimes overlapped near this wavenumber resulting in a strong Raman mode at room pressure). The mode near 235 cm$^{-1}$ is much stronger than that of 150 cm$^{-1}$ and in many cases only the band near 235 cm$^{-1}$ is observed. As in the case of tellurides, some ARMs due to Se nanoclusters have been found to shift with pressure with negative pressure coefficients[7,73,76] as those of bulk trigonal Se,[66,77,78] which is also a clear signature of Se-related modes. Therefore, our claim for trigonal Te precipitates being the cause of the ARMs in tellurides is also valid for selenides where Se precipitates have also been observed.

One additional question that must be clarified is how Se and Te precipitates can form in selenides and tellurides. In this context, we can comment on several factors for Se and Te segregation in chalcogenides. First of all, many selenides and tellurides are grown from the melt, where Se and Te excess during crystal growth is usually present.[12] Therefore, Se and Te nanoclusters or few-layer films can occur during crystal growth, especially near the sample edges, as proved in SnSe by positron annihilation spectroscopy.[79] It has also been proved that oxidation at ambient conditions seems to favor the presence of the ARMs.[18,21] On the other hand, it has also been shown that treatment with acid agents, like HCl, HF, and $HNO_3$ lead to formation of Te layers in CdTe, while dilution of bromine in methanol removes the Te layer.[9,11] High pressure has also been shown to favor sample decomposition and thus Se and Te segregation close to or after the occurrence of structural phase transitions.[7,40] Finally, it is also well known the strong sensitivity of selenides and tellurides to light due to the large absorption coefficient of visible light in these compounds due to their small bandgap. Therefore, excessive power in the laser irradiation can cause the decomposition of the samples and the segregation of Se and Te clusters or even the melting of the samples, as already reported with common lasers.[47]

In order to show the strong sensitivity of some tellurides to laser light during Raman scattering measurements and how ARMs in tellurides may occur upon excessive laser power irradiation, we have plotted in **Figure 3** the unpolarized RS of trigonal Te and other three tellurides ($\alpha$-$As_2Te_3$, $Bi_2Te_3$ and $SnBi_2Te_4$) for different laser powers. It must be noted that excitation was performed with laser light (632.8 nm) having an energy (1.98 eV) well above the bandgap of all these compounds that are small-bandgap materials. The RS of trigonal Te (**Figure 3a**) shows the main three Raman bands of Te at all laser powers. On the other hand, the RS of $\alpha$-$As_2Te_3$ and $Bi_2Te_3$ excited with low power are similar to those already published[23,80,81] and the wavenumbers of the experimental modes match with those theoretically calculated (see bottom marks in **Figure 3b and 3c**). Finally, the RS of $SnBi_2Te_4$ is here reported for the first time to our knowledge and the Raman-active modes match those theoretically calculated (see bottom marks in **Figure 3d**). Note that the RS of SnBi$_2$Te$_4$ can be nicely compared to that recently published for isostructural SnSb$_2$Te$_4$.[40][39]

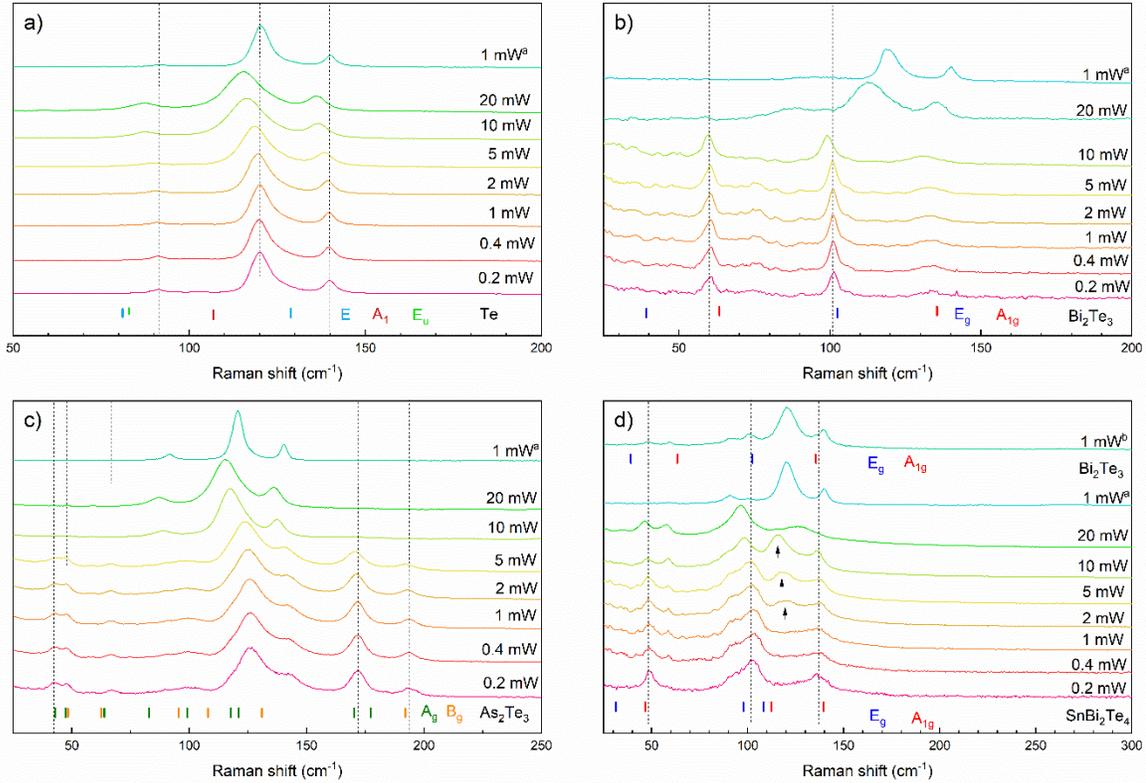

**Figure 3.** Unpolarized Raman spectra of Te (a), Bi$_2$Te$_3$ (b), $\alpha$-As$_2$Te$_3$ (c), and SnBi$_2$Te$_4$ (d) for different laser powers. Bottom black (red) tick marks show the calculated Raman-active (IR-active) TO modes of the different compounds. Dashed lines have been added as guides to the eyes. Spectra have been normalized and vertically shifted for the sake of comparison and clarity.

The Raman-active modes of Te redshift and broaden as the laser power is increased above 1 mW due to heating of the sample by excessive laser power during Raman scattering measurements (**Figure 3a**). No signal of oxidation of the Te samples due to the formation of TeO$_2$ is observed up to 20 mW. In fact, RS excited with 1 mW laser power on the same spot of the sample previously heated once the sample is thermalized (see top RS in **Figure 3a**) shows again the same features observed in RS obtained with low powers (see bottom RS in **Figure 3a**). A similar result is observed in GaGeTe heated by laser (see **Figure S4** in the SI). In this case, a redshift is observed for laser powers above 2 mW and no signal of Te or $TeO_2$ precipitates are observed even up to 20 mW, thus evidencing the high stability of the $R\bar{3}m$ phase of this compound under laser irradiation. In fact, RS obtained with 1 mW power on the same spot previously heated (see top RS in **Figure S4**) shows again the same features observed during RS measurements at low powers (see bottom RS in **Figure S4**). Therefore, we can conclude that the ARMs of GaGeTe shown in **Figure 1** do not come from laser heating or from surface oxidation, but from surface or inner Te layers formed likely bulk crystal growth.

A different scenario is found in $As_2Te_3$. Its RS shows, even at the smallest laser power, the two main ARMs around 126 and 142 $cm^{-1}$. They were previously reported and tentatively attributed to Te-related modes.[23] All the Raman bands of $As_2Te_3$ show a redshift on increasing laser power above 1 mW due to sample heating by excessive laser power; however, a change in the RS can be observed for powers above 5 mW. Under these conditions, the Raman bands of $As_2Te_3$ dissappear and only the ARS with the bands of trigonal Te are observed. Even the $E^1$(TO) mode close to 92 $cm^{-1}$ and the second-order Raman mode of trigonal Te above 260 $cm^{-1}$ are observed. In fact, the RS excited with 1 mW laser power on the damaged region due to excessive heating after proper thermalization shows the three first-order Raman peaks of trigonal Te located at 92, 121 and 140 $cm^{-1}$; i.e. at similar frequencies to those measured in pure bulk trigonal Te. Moreover, the FWHM of the three peaks are 4.9, 5.2 and 3.1 $cm^{-1}$. These values of the wavenumbers and linewidths indicate that these ARMs in burned $As_2Te_3$ are closer to those of pure Te than to those shown by GaTe and GaGeTe in **Figure 1**. These results indicate that $\alpha$-$As_2Te_3$ decomposes for powers above 5 mW and that excessive laser heating can lead to big Te clusters that are almost bulk Te. Note that no signal of the $E^1$(LO) mode is observed in the RS of burned $As_2Te_3$, unlike in GaTe and GaGeTe. This is likely due to the validity of Raman selection rules in Te layers of rather large thickness or Te precipitates of rather large size formed at the surface of the burned sample.

In $Bi_2Te_3$, Raman peaks redshift with increasing laser power above 5 mW due to sample heating and the shift of Raman peaks occurs at a similar rate as in GaGeTe and at a much smaller rate than in $As_2Te_3$. The different shift rate is likely due to a better thermal conductivity in $Bi_2Te_3$ and GaGeTe than in $As_2Te_3$ that has a larger bandgap than the formers. ARMs corresponding to trigonal Te were found only when excited with 20 mW laser power, thus evidencing a lower stability of the $R\bar{3}m$ phase of this compound under laser irradiation than GaGeTe but higher stability than $As_2Te_3$. Again, the RS excited with 1 mW laser power over the damaged sample after proper thermalization (top RS in **Figure 3b**) shows the two main Raman modes of pure trigonal Te located near 120 and 140 $cm^{-1}$; i.e. at similar wavenumbers to those measured in bulk Te. It must be noted that other rather stable tellurides seem to be $Sb_2Te_3$, $MoTe_2$ and $WTe_2$, in which trigonal Te modes have not been observed to our knowledge for any laser power irradiation or for different sample thicknesses in the literature.

As a final example of tellurides we show the RS of $SnBi_2Te_4$ (**Figure 3d**), a compound that intrinsically shows a cation disorder, as well as $SnSb_2Te_4$, even when synthesized by different routes.[82] All Raman bands redshift with increasing laser power above 1 mW due to sample heating. In fact, a change in the RS can be observed when laser power goes above 2 mW; i.e. at smaller powers than $As_2Te_3$. A broad Raman band close to 120 $cm^{-1}$ appears that seems to correspond to trigonal Te; however, on further increasing pressure to full power we observed strong changes and the ARMs of tellurides are clearly shown at 20 mW plus some remnant low-frequency modes of the original sample. The RS excited with 1 mW on burned regions after proper thermalization but in two different sample locations show slightly different results. In some regions, the RS only shows the main bands of trigonal Te near 122 and 140 cm$^{-1}$ (RS of 1mW$^a$ in **Figure 3d**), like in As$_2$Te$_3$ and Bi$_2$Te$_3$, while in other regions the ARMs corresponding to Te are observed together with a band near 50 cm$^{-1}$ that seems to correspond to original SnBi$_2$Te$_4$ and with two bands near 60 and 100 cm$^{-1}$ that can be attributed to Bi$_2$Te$_3$ (RS of 1mW$^b$ in **Figure 3d**). In summary, our RS provide evidence of both Te segregation and decomposition of SnBi$_2$Te$_4$ into its parent compounds cubic SnTe and Bi$_2$Te$_3$ with increasing laser irradiation. Since cubic SnTe does not show Raman modes only those of Bi$_2$Te$_3$ are observed. This result is in agreement with the pressure-induced decomposition already observed in SnSb$_2$Te$_4$ upon compression.[40]

Finally, to further substantiate the idea of Se precipitates in selenides we have plotted in **Figure S5** the RS of two selenides (rhombohedral $R\bar{3}m$ Bi$_2$Se$_3$ and orthorhombic *Pnma* Sb$_2$Se$_3$) for different laser powers. Bi$_2$Se$_3$ shows a RS that is similar to that already reported[83] and is very stable to laser irradiation, as rhombohedral $R\bar{3}m$ Bi$_2$Te$_3$, with no signs of Se precipitates for any laser power, except in a region where a broad band near 250 cm$^{-1}$ could be possibly attributed to Se. On the contrary, Sb$_2$Se$_3$ is rather sensitive to laser irradiation. At low power, the RS of the orthorhombic *Pnma* phase is observed in good agreement with the literature;[84-87] however, a complete different RS is observed above 5 mW (see asterisk marks) together with some low-wavenumber peaks of the original *Pnma* phase of Sb$_2$Se$_3$. RS excited with 1mW laser power (RS of 1mW$^a$ in **Figure S5b**) in the burned region after heat is dissipated show Raman peaks that correspond to the cubic phase of Sb$_2$O$_3$ (senarmontite);[86-88] i.e. complete oxidation of the sample at air conditions is promoted by laser heating. Curiously, RS obtained with 1mW laser power close to the burned region show an intense band above 230 cm$^{-1}$ that can be decomposed into two bands at 232 and 237 cm$^{-1}$ and a broad band between 430 and 500 cm$^{-1}$. Both the intense double mode and the broad band can be ascribed to the first-order Raman modes (A$_1$ and E$^2$) and to the second-order RS of trigonal Se, respectively.[57] Moreover, our results for Se segregation in $Sb_2Se_3$ and formation of cubic $Sb_2O_3$ at high laser powers are in agreement with recent works.[86,87] Therefore, our RS of $Sb_2Se_3$ clearly show that partial decomposition of the sample is observed and Se nanoclusters are segregated due to moderate laser heating.

We must note that the main Raman band in Se nanoclusters of selenides is always located between 230 and 240 cm$^{-1}$ and shows a very small Raman shift with respect to bulk values. This is in clear contrast with the strong Raman shift of the $A^1$ mode in Te nanoclusters with respect to bulk values. The small Raman shift in Se nanoclusters with respect to the bulk is in agreement with recently reported RS in few-layer Se sheets[89,90] and facilitates the identification of Se precipitates in selenides in comparison to Te precipitates in tellurides. In summary, our RS in selenides and tellurides with different excitation laser powers clearly show that care must be taken when exciting these many of these compounds with laser powers higher than 1 mW. Laser powers below this one must be used in order to avoid sample damage that can lead to partial or total decomposition of the chalcogenides and to segregation of Se and Te precipitates.

In conclusion, anomalous Raman bands of rather high intensity and linewidth have been observed in Te-based binary and ternary chalcogenides, being the two strongest bands between 119 and 145 cm$^{-1}$. On the light of the results of Raman scattering measurements on telluride bulk materials and thin films, we have proposed a very reasonable explanation for the origin of the anomalous Raman modes in tellurides. We consider that they come from the presence of Te clusters or precipitates either in form of layers or grains (typically of nanometric size unless there is a strong damage of the sample). They can be segregated at the first atomic layers of the sample surface but can be also formed inside the sample. Such segregation is usually found in bulk and 2D Te-based chalcogenides. Additional sources for the formation of pure Te precipitates can be oxidation, compression, and laser irradiation; i.e. processes that alter the delicate equilibrium of the stability of tellurides at ambient conditions. We have also shown that a similar situation occurs for some Se-based chalcogenides where Se precipitates have been also found. Therefore, special care must be taken when performing Raman characterization of selenides and tellurides, especially in 2D materials and small samples with low thermal conductivity, where thermal radiation cannot be efficiently dissipated. In those cases, very low excitation powers below 1 mW are recommended for Raman scattering measurements. We hope the present work will help interpreting the RS in selenides and tellurides in which the significance of the laser power must be taken into account for an accurate and proper Raman characterization of these light-sensitive materials.

ASSOCIATED CONTENT

**Supporting Information**.

The Supporting Information is available free of charge at:

Experimental and theoretical methods, Polarized and unpolarized Raman spectra of GaGeTe, phonon dispersion curves of GaTe and GaGeTe, Polarized and unpolarized Raman spectra of Te, unpolarized Raman spectra of GaGeTe for different laser powers, unpolarized Raman spectra of $Bi_2Se_3$ and $Sb_2Se_3$ for different laser powers. (PDF)

AUTHOR INFORMATION

**Corresponding Author**

**Francisco Javier Manjón** − *Instituto de Diseño para la Fabricación y Producción Automatizada, MALTA Consolider Team, Universitat Politècnica de València, 46022 Valencia, Spain*; ORCID: orcid.org/0000-0002-3926-1705

Email: fjmanjon@fis.upv.es

**Authors**

**Samuel Gallego-Parra** − *Instituto de Diseño para la Fabricación y Producción Automatizada, MALTA Consolider Team, Universitat Politècnica de València, 46022 Valencia, Spain;* orcid.org/0000-0001-6516-4303

**Plácida Rodríguez-Hernández** − *Departamento de Física, Instituto de Materiales y Nanotecnología, MALTA Consolider Team, Universidad de La Laguna, 38205 Tenerife, Spain;* orcid.org/0000-0002-4148-6516

**Alfonso Muñoz** − *Departamento de Física, Instituto de Materiales y Nanotecnología, MALTA Consolider Team, Universidad de La Laguna, 38205 Tenerife, Spain;* orcid.org/0000-0003-3347-6518

**Cestmir Drasar** – *Faculty of Chemical Technology, University of Pardubice, Pardubice 532 10, Czech Republic;* orcid.org/0000-0002-5645-5683

**Vicente Muñoz-Sanjosé** – *Departamento de Física Aplicada i Electromagnetismo, Universitat de València, 46100 Burjassot, Spain;* orcid.org/0000-0002-3482-6957

**Oliver Oeckler** – *Institut für Mineralogie, Kristallographie und Materialwissenschaft, Universitat Leipzig, Germany;* orcid.org/0000-0003-0149-7066

**Notes**

The authors declare no competing financial interests.

ACKNOWLEDGMENT

This work has been performed under financial support from the Spanish Ministry of Science, Innovation and Universities, the Spanish Research Agency (AEI), the European Fund for Regional Development (FEDER) under grants TEC2017-85912-C2-2, RED2018-102612-T (MALTA-CONSOLIDER TEAM network) and PID2019-106383 GB-C42/C43. We also thank financial support from Generalitat Valenciana under grant PROMETEO/2018/123 (EFIMAT). Authors also acknowledge Dr. Philipp Urban for performing the synthesis of $SnBi_2Te_4$.

Supporting Information for

# Anomalous Raman Modes in Tellurides

F. J. Manjón,[1,*] S. Gallego-Parra,[1] P. Rodríguez-Hernández,[2] A. Muñoz,[2]

C. Drasar,[3] V. Muñoz-Sanjosé,[4] and O. Oeckler[5]

[1] Instituto de Diseño para la Fabricación y Producción Automatizada, MALTA Consolider Team, Universitat Politècnica de València, 46022 Valencia, Spain

[2] Departamento de Física, Instituto de Materiales y Nanotecnología, MALTA Consolider Team, Universidad de La Laguna, 38205 Tenerife, Spain

[3] Faculty of Chemical Technology, University of Pardubice, Pardubice 532 10, Czech Republic

[4] Departamento de Física Aplicada y Electromagnetismo, Universitat de València, 46100 Burjassot, Spain

[5] Institut für Mineralogie, Kristallographie und Materialwissenschaft, Universität Leipzig, Germany

* corresponding author: fjmanjon@fis.upv.es

**Experimental and theoretical methods**

Bulk single crystals and polycrystalline samples with thickness larger than 50 mm were used in this work. Monoclinic GaTe, orthorhombic $Sb_2Se_3$ and rhombohedral GaGeTe, $Bi_2Se_3$ and $Bi_2Te_3$ were grown by the Bridgman method **[1,2,3]**, $SnBi_2Te_4$ was prepared in a silica glass ampoule **[4]**, while monoclinic $\alpha$-$As_2Te_3$ **[5]** and trigonal Te samples were commercially acquired.

Raman scattering measurements on all samples were performed in backscattering geometry at ambient conditions in air using a 50x LWD objective coupled to a Horiba Jobin Yvon HR800 UV microspectrometer with a thermoelectrically cooled CCD camera. Raman signal was excited with a vertically-polarized HeNe laser (632.8 nm) of 20 mW power. Neutral density filters were used to excite samples from 20 mW down to 0.2 mW. Low powers were employed to avoid damage to the samples while high powers were used to intentionally damage the samples. A 1200 lines/mm grating provided an experimental resolution of 1.6 $cm^{-1}$. Ultra-low frequency measurements down to 10 $cm^{-1}$ were carried out with a set of volume Bragg grating filters for the 632.8 nm line. Polarization in different configurations were achieved by rotating the sample with respect to the incident polarized laser light and by placing horizontal and vertical polarizers prior to the entrance of the Raman signal into the spectrometer. Analysis of Raman spectra has been performed by taking the 520 $cm^{-1}$ Raman line of Si as a reference, by substracting the corresponding background and by fitting Voigt profiles to the Raman peaks in which the Gaussian width is fixed to the experimental resolution.

*Ab initio* theoretical calculations were carried out within the framework of density functional theory (DFT)**[6]** with the Vienna Ab-initio Simulation Package (VASP)**[7]**, using the pseudopotential method and the projector augmented waves (PAW) scheme **[8,9]**. In this work, the generalized gradient approximation (GGA) with the Perdew-Burke-Ernzerhof (PBE) parametrization extended to the solid state (PBEsol) was used for the exchange and correlation energy **[10]**. Lattice-dynamical properties were obtained for the $\Gamma$-point using the direct-force constant approach **[11]**.

**Figure S1.** Polarized and unpolarized Raman spectra of GaGeTe showing the anomalous Raman modes. Spectra have been normalized and vertically shifted for the sake of clarity.

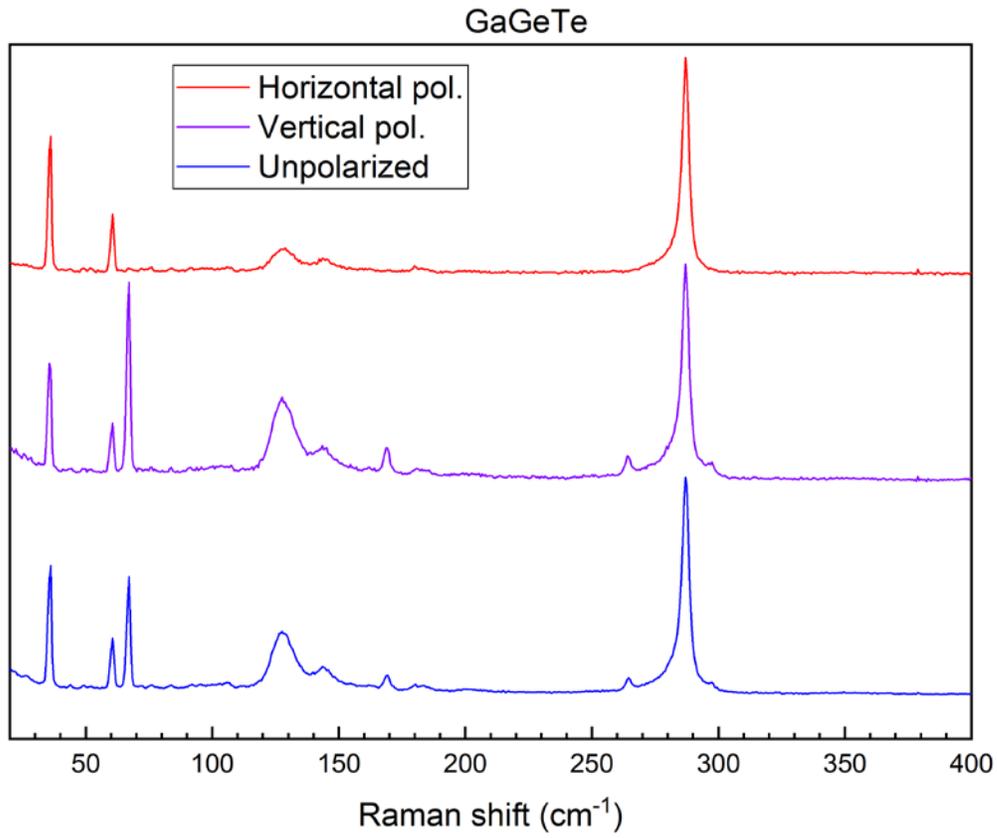

**Figure S2.** Phonon dispersion curves of monoclinic GaTe (a) and rhombohedral GaGeTe (b).

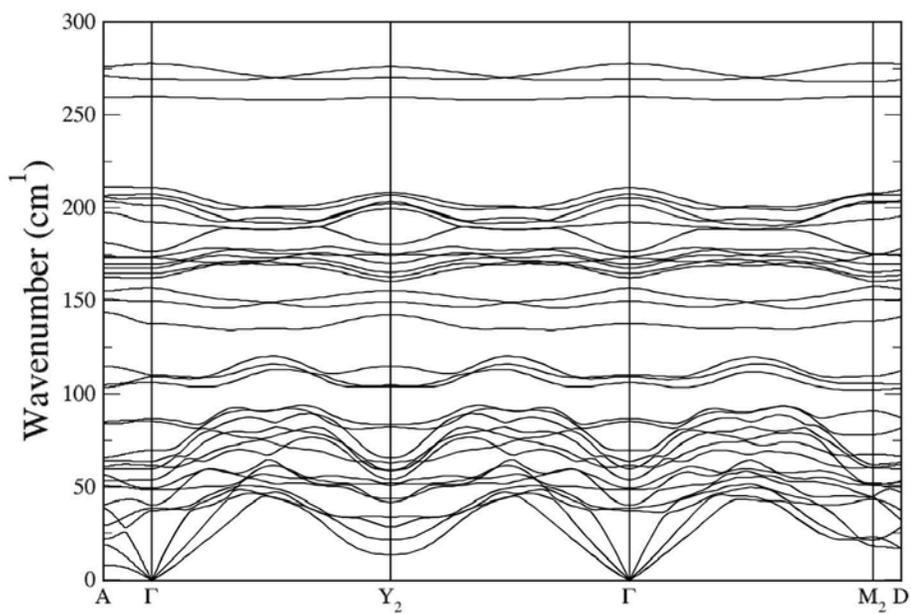

(a)

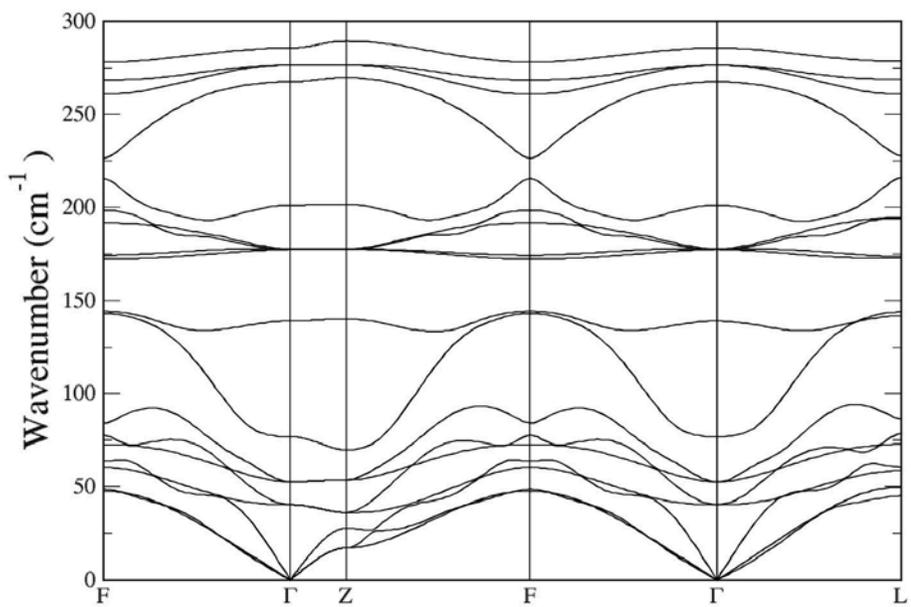

(b)

**Figure S3.** Raman spectra of trigonal Te in backscattering geometry. (a) With laser polarization parallel to the rods; i.e. the c axis (E ∥ c). (b) With laser polarization perpendicular to the rods (E ⊥ c). (c) With laser polarization at 45° of the rods (E@45°). Unpolarized as well as parallel (vertical pol.) and cross (horizontal pol.) polarized measurements were performed. Bottom black (red) marks show the calculated Raman-active (IR-active) TO modes of Te. Note that the two E modes of Te are also IR-active. Spectra have been normalized and vertically shifted for the sake of clarity.

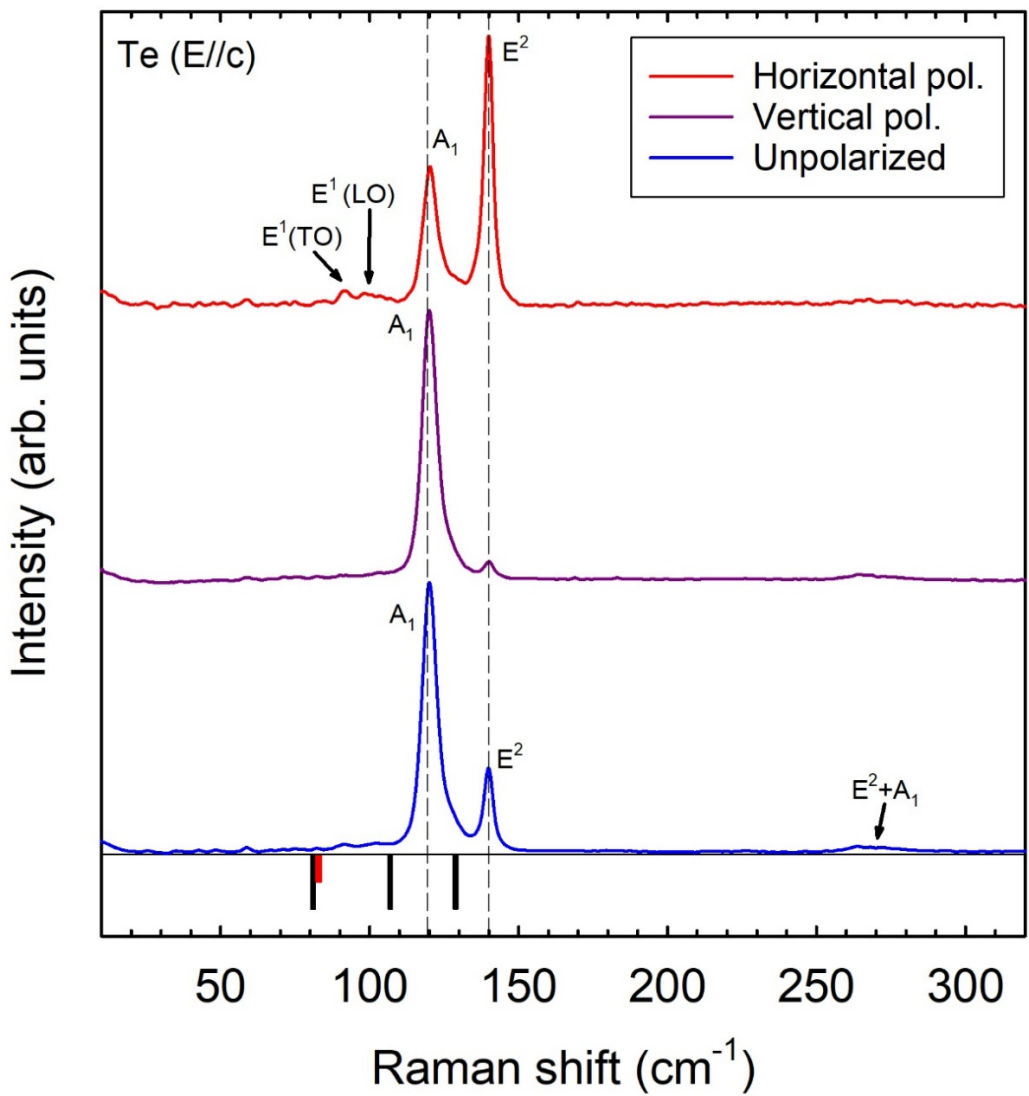

(a)

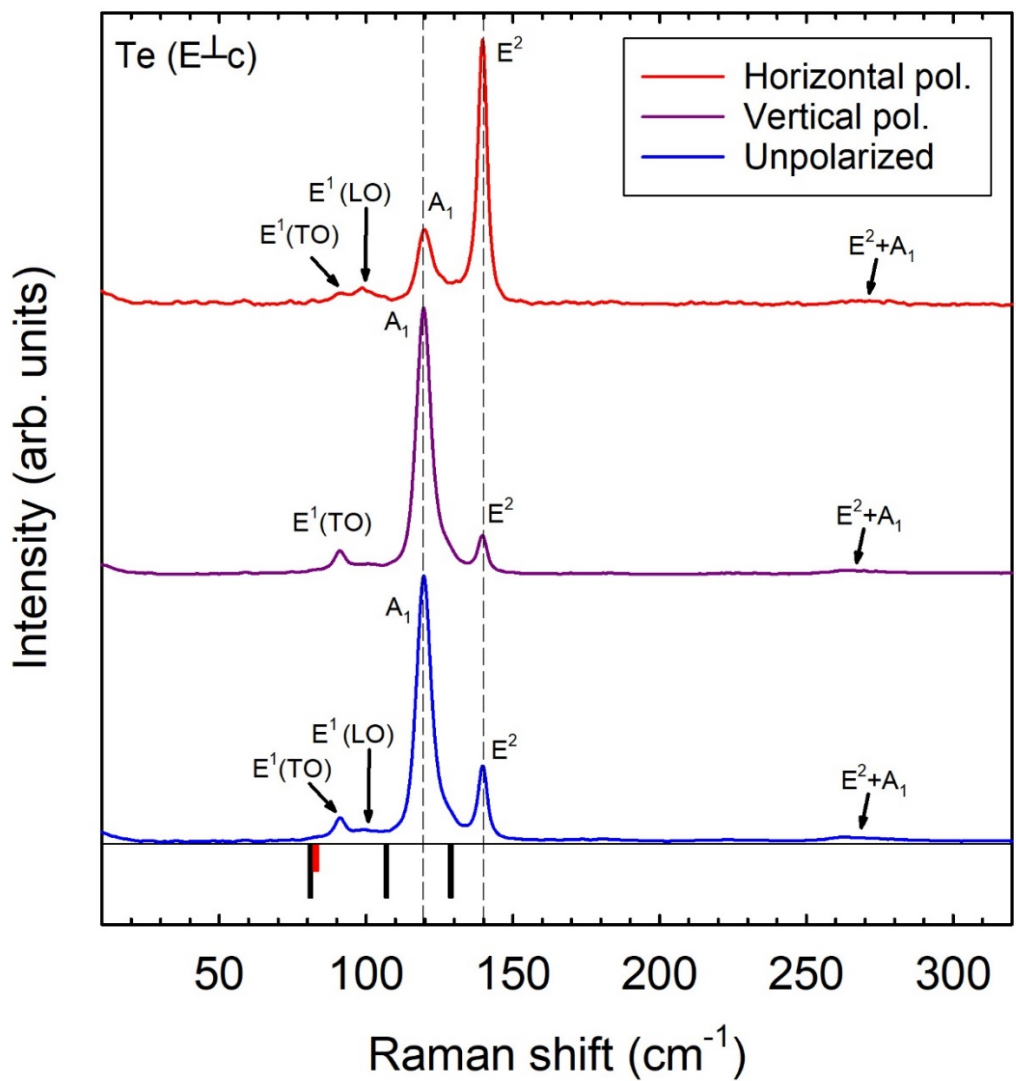

(b)

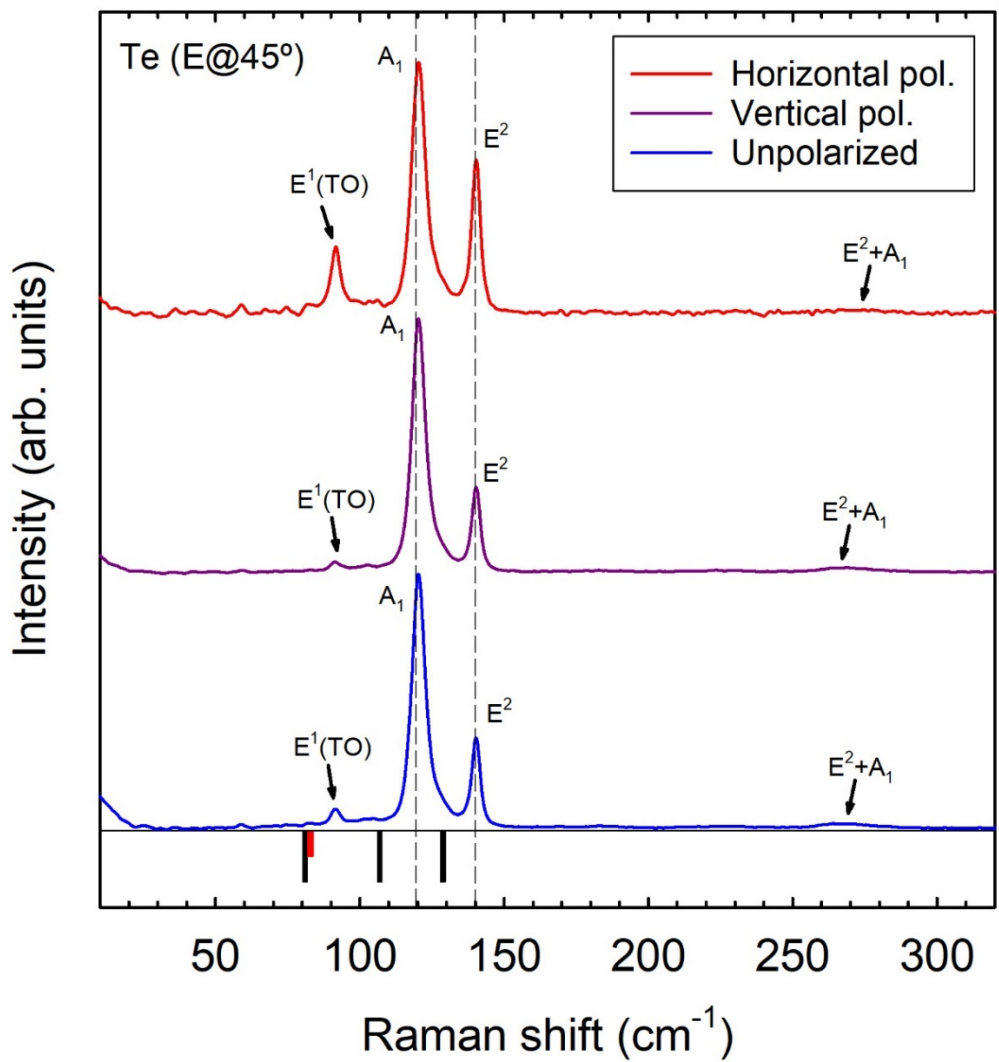

(c)

**Figure S4.** Unpolarized Raman spectra of GaGeTe for different laser powers. Spectra have been normalized and vertically shifted for the sake of clarity.

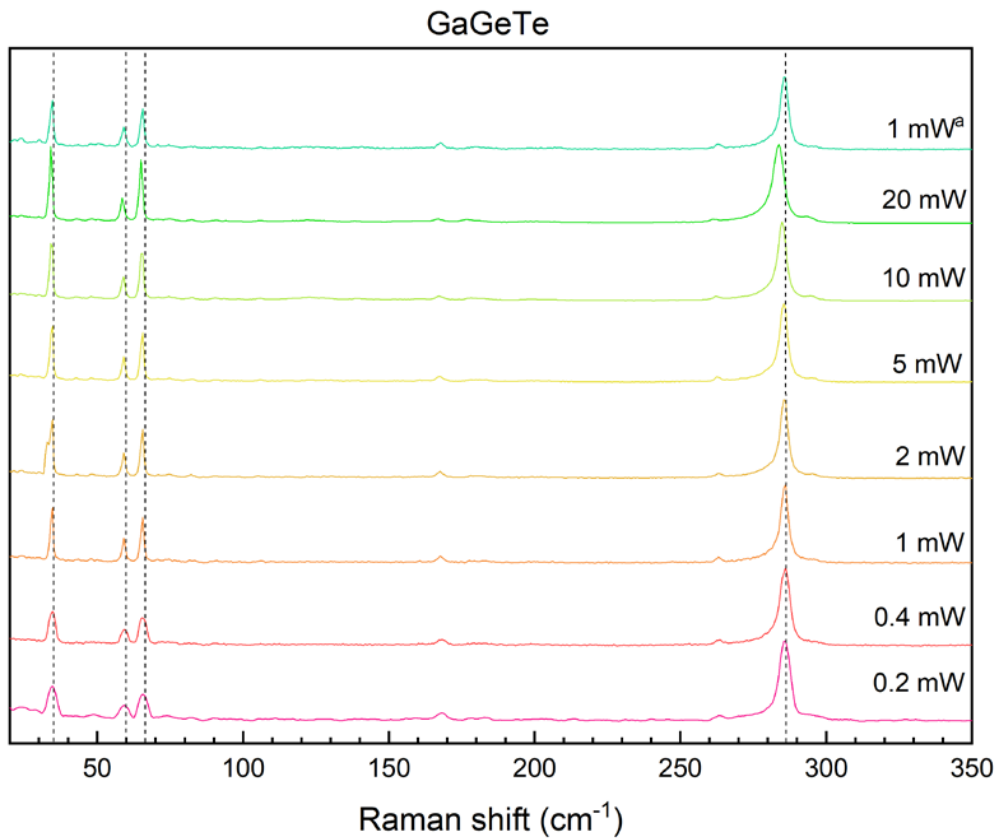

**Figure S5.** Unpolarized Raman spectra of Bi$_2$Se$_3$ (a) and Sb$_2$Se$_3$ (b) with different laser powers. Spectra have been normalized and vertically shifted for the sake of clarity.

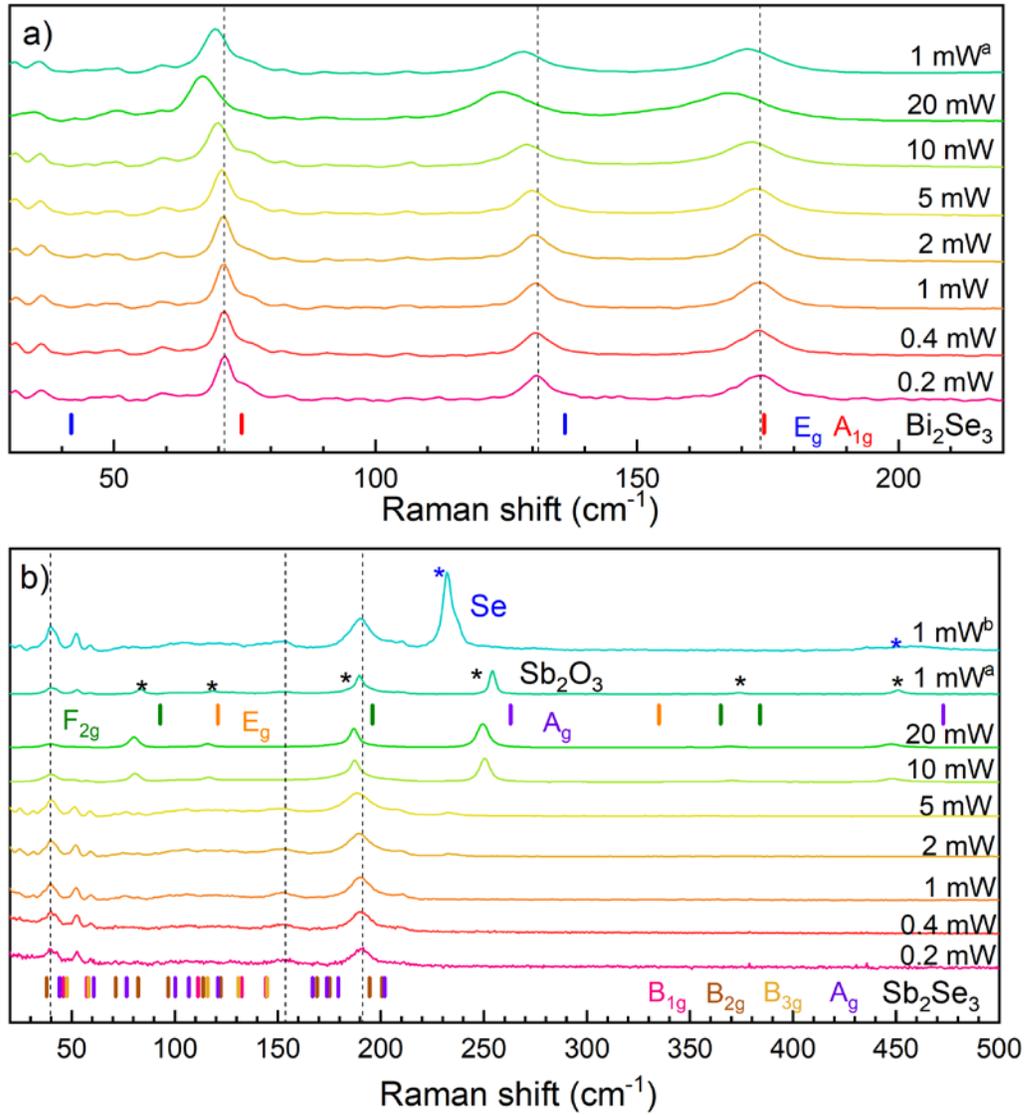